\documentclass[a4paper]{report}
\usepackage[utf8]{inputenc}
\usepackage[T1]{fontenc}
\usepackage{RJournal}
\usepackage{amsmath,amssymb,array}
\usepackage{booktabs}
\usepackage{algorithm}
\usepackage[noend]{algpseudocode}

\begin{document}

%% do not edit, for illustration only
\sectionhead{Contributed research article}
\volume{XX}
\volnumber{YY}
\year{20ZZ}
\month{AAAA}

%% replace RJtemplate with your article
\begin{article}
  % !TeX root = RJwrapper.tex
\title{\pkg{fc}: A Package for Generalized Function Composition Using Standard Evaluation}
\author{by Xiaofei Wang and Michael John Kane}
% packages to examine
% 1. functional (allows for function compositions, but assumes only one argument? also not updated since 2014)
\maketitle

\abstract{
In this article, we present a new R package \pkg{fc} that provides a streamlined, standard evaluation-based approach to function composition. Using \pkg{fc}, a sequence of functions can be composed together such that returned objects from composed functions are used as intermediate values directly passed to the next function. Unlike with \CRANpkg{magrittr} and \CRANpkg{purrr}, no intermediate values need to be stored. When benchmarked, functions composed using \pkg{fc} achieve favorable runtimes in comparison to other implementations.
%Currently, there are a number of ways to apply a sequence of functions to a data object in R. For example, the sequence of functions could be specified via a pipeline syntax using \CRANpkg{magrittr}, yielding a wrapper function containing a list of functions to be applied. At runtime, the wrapper function proceeds through the function list in order, storing return objects from each function as intermediate values that get passed to the next function. In this paper, we present a new R package \pkg{fc} that provides an alternate approach using function composition, wherein the sequence of functions is composed together so that no intermediate values need to be stored. Furthermore, the package enables a similar pipeline syntax and achieves favorable runtimes compared to other implementations. 
}

\section{Introduction}

Coding in R often reduces to running a sequence of operations in order. For instance, as part of exploratory data analysis, one might subset the first 50 rows from a data frame and then examine numerical summaries of the subset. This could be accomplished using base R for a given data frame:
\begin{example}
  summary(head(iris, 50))
\end{example} 
Via \CRANpkg{magrittr} \citep{magrittr}, the use of a pipe operator improves readability, enabling the user to read the sequence of functions in the order of application from left to right:
\begin{example}
  iris %>% head(50) %>% summary()
\end{example} 
%With proper function composition, the sequence of operations %could be encapsulated within a portable, new function. The %newly-created function could then be applied to any new datasets %efficiently at runtime, without the overhead of recreating %intermediate steps. This is not possible with the use of %\CRANpkg{magrittr}, which mandates that each pipeline begins %with a data object and then applies each successive function to %modify the data object.
The analysis pipeline, defined by \code{head(50) \%>\% summary()}, is a sequence of functions that are
run in succession. The first function uses the data object on the far left as input. 
%Each function uses, as input, either data or the output of another function, which is stored as an intermediate variable, usually named ``.'' by convention.
Each successive function is run to completion, using the returned object from the function left of each pipe as input to the function on the right.

Over the past few years, the pipe operator (\verb|%>%|) has become a popular
means for chaining together a sequence of functions, particularly where a long chain of commands needs to be run, as often is the case with data wrangling 
%say for transforming an unstructured set of data to a structured one, or where a standard set of transformations is performed on a structured dataset, such as those defined in 
\citep{wickham2016r}. A good deal of this popularity comes from the economy with which pipelines of functions are created. Construction of pipes often relies heavily on {\em non-standard evaluation} (NSE) \citep{Wickham2014}, where parsed expressions must be manipulated before their evaluation to ensure correct execution. For example, in the expression \verb|iris %>% head(50) %>% summary()|, both the \code{head()} and \code{summary()} functions are altered so that the \code{iris} and output of \code{head()} respectively are inserted as the first function arguments. This {\em dialect} of \R \ has seen widespread adoption due to its conciseness, enabling the user to quickly define sequences of operations to perform on data. 
%However, it does not provide fundamental capabilities that were previously beyond those of \R. 

In this paper, we present a standard evaluation approach to chaining functions via function composition; the return of the first function becomes the argument to a second function and so on, obviating the need for storing intermediate variables and function lists. This approach, which optionally includes a slightly different pipeline syntax, is implemented in a new R package \pkg{fc}. 
%This approach only works for function composition; the pipeline must be defined separately from its use. 

In the ensuing sections, we begin with a review of existing work in function chaining and function pipelines. We consider the fundamental differences between how pipe-forward operators are used in \pkg{magrittr} and \pkg{fc}, and show how function composition is achieved using \pkg{fc}. We then describe the underlying implementation of the \pkg{fc} package. We additionally present a number of examples that show the diverse applications of \pkg{fc}, including compatibility with packages like \pkg{dplyr} \citep{dplyr}; we show that runtimes are comparable and at times superior to implementations using other packages. 

%In this article, we introduce a new \pkg{fc} package that %performs function composition while optionally supporting the %syntactic sugar of \CRANpkg{magrittr}-esque pipes. Unlike %\CRANpkg{magrittr}, \pkg{fc} uses standard evaluation, which %contributes to improved runtimes. 

%Another package that specializes in function composition in R %albeit without prioritizing readability is \CRANpkg{purrr} %\citep{purrr}. We will show that both \CRANpkg{purrr} and %\CRANpkg{magrittr} tend to run more slowly compared to \pkg{fc} %when applying a function sequence on a data object. In the case %of \CRANpkg{magrittr}, this is likely due to not actually %implementing function composition. Furthermore, both adopt a %non-standard evaluation approach, which results in some overhead %in parsing the code.

%The \CRANpkg{magrittr} packages is currently the most popular %implementation
%of the pipe-forward operator - partially because of its %inclusion in the ``tidyverse'' \citep{Ross2017} and
%For the rest of this paper, references to the current %\code{\%>\%}
%operator assume this implementation.

\section{Background}

To our knowledge, \CRANpkg{purrr} \citep{purrr} and \CRANpkg{magrittr} are the only R packages under active development that provide procedural composition of functions -- namely, both packages are capable of encapsulating a specified sequence of functions within a single function. Within \CRANpkg{purrr}, such encapsulating functions can be created via \code{compose()} and within \CRANpkg{magrittr}, they are created using a pipeline with `.' in place of the data object that ordinarily precedes the first pipe operator. Despite differences in syntax, their output function bodies are actually the same -- a list of functions stored in the encapsulating function in run order. Of the two packages, \CRANpkg{magrittr} is by far the more commonly-used package due to its convenient pipeline syntax. Below, we give an overview of the history of pipes in R programming and then describe their use in \CRANpkg{magrittr}. 

\subsection{A History of Pipes}

The characterization of the \verb|%>%| operator as a ``pipe'' may be attributed to the
\pkg{pipeR} package \citep{pipeR}; its associated syntax is inspired by Unix
pipes. It may be noted, though, that \verb|>| is not actually the pipe operator
in Unix; it is the {\em indirection operator} used to route the output
of a pipe to a file, including \code{stdin}, \code{stdout}, and \code{stderr}. The Unix pipe
operator is denoted as \verb|||. Nonetheless, the convention
of using \verb|>| as a pipe is not without precedence; for example, the
F\# programming language denotes \code{|>} as the pipe-forward
operator. 

Along with syntactic differences, R pipes differs from Unix pipes in their design.
A Unix pipe defines a sequence of applications run as different processes
where the output of each application is formatted text that can be
read into the next application in the ``pipeline.'' Text is outputted
in contiguous blocks or ``chunks'' using files.
Applications run concurrently, processing input as they become
available, thereby managing memory usage (chunks are stored in the
file cache and may be much smaller than the input) as well as load
balancing (applications are bottlenecked by the one with
the lowest throughput). Pipeline concurrency within R is supported in the \CRANpkg{iotools} package
\citep{iotools} but its use is primarily focused on chunked file processing and it does not support a pipe-forward operator.

\subsection{\pkg{magrittr} Pipes and Non-Standard Evaluation}

When used with functions, \pkg{magrittr}'s pipe-forward is an infix operator collecting the sequence of functions to be applied to an input data object. NSE alters each function call in the pipeline so that it can be evaluated in the standard way and its return value can be (1) returned if it is the last function call or (2) sent to the next function in the sequence of functions defined by the pipeline. Continuing with our opening example, we can create a function using \CRANpkg{magrittr} as follows:
\begin{example}
  head_summary <- . %>% head(50) %>% summary()
\end{example}
The new function, \code{head\_summary()}, can be examined via:
\begin{example}
  > unclass(head_summary)
  function (value)
  freduce(value, `_function_list`)
  <environment: 0x7fab1e3518c0>
\end{example}
We see that \code{head\_summary()} is a function that calls \code{freduce()} on the input and the sequence of functions stored in the \code{head\_summary()} function environment named \code{'\_function\_list'}, seen below:
\begin{example}
  > environment(head_summary)[['_function_list']]
  [[1]]
  function (.)
  head(., 50)

  [[2]]
  function (.)
  summary(.)
\end{example}
Both \code{head()} and \code{summary()} are altered in the call to pipe-forward. Each gets wrapped in a function taking `\verb|.|' as an argument; `\verb|.|` serves as the first argument in \code{head()} and \code{summary()} within the wrapper functions. 

The arguments to pipe-forward (e.g. \verb|head(50)| and \verb|summary()|) are {\em syntactically valid}; they can be parsed to create a valid expression. They are not {\em semantically valid} because the user must provide \code{x} and \code{object} arguments to \code{head()} and \code{summary()} respectively for successful execution (assuming the base R implementations of both functions). NSE turns syntactically valid but semantically invalid expressions into new syntactically and semantically valid expressions. 

Functionally, the pipe-forward provides both {\em partial multivariate function composition}, where the input arguments are the return of other functions,  as well as {\em partial multivariate function valuation}, where an argument is set to a constant (e.g. setting \verb|n| to 50 in \verb|head()|). However, we have already seen that it does not compose functions in the expected way. Otherwise, \code{head\_summary()} ought to be defined similarly to:
\begin{example}
  function(.) 
  {
     summary(head(., 50))
  }
\end{example}
Instead, \code{head\_summary()} keeps track of the sequence of functions, applying the input to the first function and the return of the first function to the first argument of the second function. The resulting output, when applied to a data object, is equal to what would be achieved using function composition.

\section{Partial Function Composition and Valuation with \pkg{fc}}

Inspired by \pkg{magrittr}, the \pkg{fc} package provides partial multivariate function composition and valuation where resulting functions are true compositions of the inputs. Composed functions send the result of one function directly to a subsequent function without intermediate values. Code analysis finds unbound variables (variable names with no associated value) and includes them as variables in returned function.
%, thereby addressing the case where an intermediate value is needed by multiple inputs. 

For convenience, a pipe-forward operator is included in the package and can be thought of as an application of the more general functionality. The resulting implementations from either of these function-composing approaches tend to be more readable and easier to debug compared to that of \pkg{magrittr}.

\subsection{The fc() function}

The primary function in the \pkg{fc} package is \verb|fc()|, which simultaneously allows for partial function composition and valuation. The first argument to \verb|fc()| could either be an existing function in the environment or an anonymous function. Any following arguments must be named arguments to this function. For example, we can create a new function that uses partial function valuation to display the first 50 rows of a dataset with:
\begin{example}
  head50 <- fc(head, n=50)
\end{example}
The return function has a single argument \verb|x|, inherited from the \code{head()} function. The function \verb|head50()| consists of:
\begin{example}
  function (x) 
  {
    head(x, n = 50)
  }
\end{example}

In order to perform function composition, multiple \verb|fc()| calls could be used in a nested manner:
\begin{example}
  summary50 <- fc(summary, object=fc(head, n = 50)(object))
\end{example}
Or, \code{head50()} could be defined separately, and we could rewrite:
\begin{example}
  summary50 <- fc(summary, object=head50(object))
\end{example}

The signature of the returned function consists of: (1) parameters in the formals of the function passed to \code{fc()} that are not assigned as well as (2) the unbound symbols in the expressions passed as arguments to \code{fc()}. As a result, the signature of the return function can be specified differently than that of the function it takes as its first argument. For example, if we wanted the signature of \code{summary50()} to be a single variable called \code{x}, then we could specify:
\begin{example}
  summary50 <- fc(summary, object=head50(x))
\end{example}
with function definition given by:
\begin{example}
  function (x)
  {
    summary(object = head50(x))
  }
\end{example}

All arguments to \code{fc()}, except the first, must be named. 
%While this restriction could be relaxed in subsequent implementations, 
This decision was made to minimize confusion that may occur when associating expressions with arguments in the mixed named/positional argument setting. The \code{fc()} function also does not promote unbound argument-function variables to arguments of the returned function. This is because some functions distinguish between a default argument and an argument with a default argument that is not passed an input at run-time.

\subsection{Default and Passed Arguments}

One function making the distinction between default and passed arguments is \code{matrix()}, whose signature is
\begin{example}
  > matrix
  function (data = NA, nrow = 1, ncol = 1, byrow = FALSE, dimnames = NULL)
  ...
\end{example}
If unassigned arguments were promoted to input function arguments in a hypothetical \code{fc\_bad()} function, then
\begin{example}
  matrix_bad <- fc_bad(matrix, ncol = 3)
\end{example}
would result in a function
\begin{example}
  function(data = NA, nrow = 1, byrow = FALSE, dimnames = NULL)
  {
    matrix(data = data, nrow = nrow, ncol = 3, byrow = byrow, 
      dimnames = dimnames)
  }
\end{example}
and the following call would result in incorrect behavior, which does not produce either a warning or an error:
\begin{example}
  > matrix_bad(rnorm(9))
           [,1]     [,2]       [,3]
  [1,] 1.738805 1.232176 0.01928793
\end{example}
This is different than the 3 $\times$ 3 matrix we would expect to see from the direct call \code{matrix(rnorm(9), ncol = 3)}. The difference comes from the fact that the direct call detects that \code{nrow} is missing and calculates the correct number of rows despite the fact that the default number of rows is 1 in the signature. The hypothetical \code{fc\_bad()} function fails because the default values in the argument-function were taken as the defaults in the return function. As a result, when \code{fc(matrix\_bad, rnorm(9))} is called, with the promoted defaults, the \code{matrix()} function is called with the default \code{nrow = 1}. Instead, we need to specify that the \code{data} argument is a return-function parameter and the \code{ncol} argument is 3:
\begin{example}
  > matrix_3_col <- fc(matrix, data=data, ncol=3)
  > matrix_3_col
  function (data)
  {
    matrix(data = data, ncol = 3)
  }
\end{example}
The composed function then operates as expected:
\begin{example}
  > matrix_3_col(rnorm(9))
              [,1]      [,2]         [,3]
  [1,]  0.04834349 0.4315403  0.002819384
  [2,]  1.39077410 1.1928053  0.980652418
  [3,] -0.38902263 1.1849566 -1.239682276
\end{example}

\subsection{The Pipe-forward Operator}

A pipe-forward operator is included in the package and can be regarded as a special case of functionality provided by the \code{fc()} function. This is because both the left- and right-hand-side functions provided to the pipe-forward operator take a single input (after appropriate argument modification). 
Therefore, the implementation of the pipe-forward operator is simply the composition of two functions, provided by the \code{fc()} function. As a result, contiguous, syntactically meaningful subsets of a valid \code{fc}-pipeline, when evaluated, result in a valid function.  
\begin{example}
  summary50 <- fc(head, n=50) %>% fc(summary)
\end{example}
Each entry in a \pkg{fc} pipeline should be a function class object. In the absence of other arguments to \verb|fc()|, the last entry of the pipeline, \verb|fc(summary)|,  produces a simple wrapper function for the \verb|summary()| function. For that reason, we can replace this last entry in the pipeline with \verb|summary|:
\begin{example}
  summary50 <- fc(head, n=50) %>% summary
\end{example}

%In contrast to \CRANpkg{magrittr} where each item in the pipeline is either a data object or a call to a function that gets applied to the data, each item in the \pkg{fc} pipeline is a \verb|fc()| function call. This retains the integrity of standard evaluation -- each item in the pipeline is proper command under R standard evaluation. 

We highlight two important distinctions between \pkg{fc}'s pipe-forward operator and that of \pkg{magrittr}. First, \pkg{fc}'s pipe-forward does not support a dataset as the left-hand-side argument. The \pkg{magrittr} package supports expressions like:
\begin{example}
  iris %>% head %>% summary
\end{example}
because the \code{head()} function is called, with \code{iris} as the input and the result is stored as the intermediate variable ``\code{.}'' by convention. This intermediate variable is then provided as the first argument to \code{summary()} in the subsequent call to the pipe-forward operator. The the pipe-forward operators have the same precedence and therefore they are parsed from left to right, according to R's grammar. However, the equivalent expression using \code{fc()}, \code{summary(head(x))}, requires the composition of \code{summary()} to \code{head()} {\em before} application to \code{iris}. Giving precedence to later steps in the pipeline is incompatible with R's parsing rules without the use of parentheses. As a result, the composition of function and the application of data to those functions must be separated, either calling the function as an intermediate variable
\begin{example}
  (head %>% summary)(iris)
\end{example}
or by creating a named function and then applying it to the data.
\begin{example}
  hsummary <- head %>% summary
  hsummary(iris)
\end{example}

The second distinction is that the \pkg{fc} package does not use the same shorthand notation for modifying multiple arguments within functions. %altering functions so that they are single-input, single-output. 
For example, in \pkg{magrittr} we might create a summary of the first half of a dataset as:
\begin{example}
  first_half_magrittr <- . %>% head(n=round(nrow(.)/2)) %>% summary
\end{example}
The \code{head()} function is altered to take half of the number of rows of the input as the argument to \code{n}, using the NSE ``\code{.}'' convention. With the \pkg{fc} package, we use \code{fc()} to explicitly create a single-input function out of \code{head()}:
\begin{example}
  first_half_fc <- fc(head, n=round(nrow(x)/2)) %>% summary
\end{example}
yielding the function:
\begin{example}
  function (object)
  {
    summary(object = internal_anon_func(object))
  }
\end{example}
where \code{internal\_anon\_func} is the result of the \code{fc} expression:
\begin{example}
  function (x)
  {
    head(x, n = round(nrow(x)/2))
  }
\end{example}

\subsection{Compatibility with Functions Having Expressions as Argument Names}

An important caveat with \code{fc()} is that it does not support using expressions as argument names. As a result, functions like \code{subset()} cannot be called as:
\begin{example}
  > fc(subset, Sepal.Length > 5)(iris)
  Error in fc(subset, Sepal.Length > 5) :
    All parameter arguments must be named.
\end{example}
However, \code{subset()} can work with named arguments:
\begin{example}
  fc(subset, subset = Sepal.Length > 5)
\end{example}

The use of expressions as function names should be used judiciously or even discouraged in since they are either executed within a specific context (for \code{subset()}, after the first argument is \code{attach()}'ed) or they require defining operators on unbound variables. It should be noted though, that this restriction does not preclude \pkg{fc}'s use with \pkg{tidyverse} packages however \citep{tidyverse}; see Example 3 in the Examples section for details. 

\section{Implementation}

The \code{fc()} function takes, as input, a function and a set of named arguments corresponding to those of the input function and whose values are expressions. Unbound variables in those expressions are used as input parameters to the resulting function that is returned by the \code{fc()} function. A high-level description of the steps taken is shown in Algorithm \ref{high_fc}.

\begin{algorithm}
    \caption{High-level Pseudocode for the \code{fc()} function}
    \label{high_fc}
    \begin{algorithmic}[1] % The number tells where the line numbering should start
        \Procedure{Function Compose}{\code{func,}...} 
          \State \code{fc\_ret\_env} $\gets$ the parent environment.
          \If {\code{func} is a call to Function Compose or anonymous function}
            \State \code{func} $\gets$ the evaluated \code{func} expression
            \State \code{fc\_ret\_env} $\gets$ a new environment
            \State \code{func} is stored in \code{fc\_ret\_env}
          \EndIf
          \If {Any \code{...} arguments are not named}
            \State return an error
          \EndIf
          \If {Any \code{...} arguments are Function Compose statements}
            \State evaluate them and keep them in \code{fc\_ret\_env}
          \EndIf
          \State \code{ret\_fun\_body} $\gets$ the name of \code{func} with parameters defined by the \code{...} arguments.
          \State \code{ret\_fun\_sig} $\gets$ the unbound arguments in each of the \code{...} expressions.
          \State Return the function defined by \code{ret\_fun\_sig}, \code{ret\_fun\_body}, and \code{ret\_fun\_env}
        \EndProcedure
    \end{algorithmic}
\end{algorithm}

The first argument may be a named function, an anonymous function, or a call to the \code{fc()} function. If it is either an anonymous function or a call to \code{fc()}, then the argument is evaluated and should result in a function. In this case, a new environment is created to hold the evaluated function. This new environment will be that of the return-function. If \code{fc()} does not need to create an anonymous function, then the return-function's environment is set to the environment returned by \code{parent.frame}, the default environment when the \code{function()} function is called. Subsequent arguments, which must be named, may be expressions, including calls to \code{fc()}, which are processed in a manner similar to that of the first argument. Subsequent arguments may not be calls to anonymous function declarations.

The signature of the return function is derived by parsing the arguments of each of the expressions supplied to the \code{...} variable. The expression tree is created using the \pkg{codetools} package \citep{codetools}, finding each of the leaves of the abstract syntax tree, that are of type \code{symbol} (corresponding to unbound variables). The unique set of symbols for each of the arguments are collected to create the return-function signature. The \code{func} and \code{...} arguments are used to generate the function body. This, along with the function signature and environment are sufficient to create the function that is returned by \code{fc()}. 

\section{Examples} % (fold)
\label{sec:examples}

In this section, we present four example usages of \pkg{fc}. In the Comparison subsection, we show the runtimes of our composed functions in these examples as well as similar implementations using \CRANpkg{magrittr} and \CRANpkg{purrr}.

\subsection{Example 1} % (fold)
\label{sub:example_1}

In a simple example involving primitive R functions only, we take the natural log of the square root of 10. This could be done using base R as follows:

\begin{example}
  log_sqrt_base <- function(x) log(x=sqrt(x))
\end{example}

We can implement the same function using \pkg{fc} in two ways, with or without pipes.

\begin{example}
  log_sqrt_fc <- fc(log, x=sqrt(x))
  log_sqrt_fc_pipe <- fc(log, x=x) %>% fc(sqrt, x=x)
\end{example}

\subsection{Example 2} % (fold)
\label{sub:example_2}
The next example involves text processing. A common web scraping task requires processing information between HTML tags found within a character string vector. This could be done in three steps: (1) using \verb|grep()| to identify entries in a vector containing information, (2) using \verb|gsub()| to extract information in those entries, and (3) cleaning up the information, say via \verb|trimws()|. A base R composed function for this task is as follows:

\begin{example}
  search_trim_base <- function(v) {
    trimws(gsub(grep(v, pattern="<[^/]*>", value=TRUE), 
    pattern=".*>(.*)<.*", replacement = "\\1"))
  }
\end{example}

Below, we show a concrete example of an input \verb|x| and the associated function output \verb|search_trim_base(x)|.

\begin{verbatim}
  x <- c("<td class = 'address'>24 Hillhouse Ave.</td>", 
         "<td class = 'city'>New Haven</td>",
         "</table>")
  search_trim_base(x)
  [1] "24 Hillhouse Ave." "New Haven"  
\end{verbatim}

We can express the same function using \pkg{fc} as follows:

\begin{example}
  search_trim_fc <- fc(trimws, 
                       x = fc(gsub, pattern=".*>(.*)<.*", 
                              replacement = "\\1", 
                              x = fc(grep, pattern="<[^/]*>", value=TRUE)(x))(x))
\end{example}
There are three calls to \verb|fc| nested together. Pipes can be used to enhance the readability:
\begin{example}
  search_trim_fc_pipe <- fc(grep, pattern="<[^/]*>", value=TRUE) %>%
    fc(gsub, pattern=".*>(.*)<.*", replacement = "\\1") %>% trimws
\end{example}

%Note that because the last function to be applied is \verb|trimws()|, applied without partial valuation, the last entry in the pipeline could either be \verb|fc(trimws)| or simply \verb|trimws|.

\subsection{Example 3} % (fold)
\label{sub:example_3}
This next example shows how \pkg{fc} can be used in conjunction with existing data wrangling packages such as \CRANpkg{dplyr}, which natively uses NSE function chaining via \CRANpkg{magrittr} pipes. Most of the NSE functions in \CRANpkg{dplyr} have standard evaluation analogues that are designated with an underscore suffix. We will use these standard evaluation functions with \pkg{fc}.

In this example, we use the \CRANpkg{nycflights13} package \citep{nycflights13} to demonstrate how to calculate group-wise numerical summaries using \CRANpkg{dplyr}. Below, we compare the usual \CRANpkg{dplyr} NSE syntax to the SE syntax supported by \pkg{fc}:

\begin{example}
  library(nycflights13)
  # using NSE, dplyr
  flights %>% group_by(tailnum) %>%
  summarize(count = n(),
            dist = mean(distance, na.rm=TRUE),
            delay = mean(arr_delay, na.rm = TRUE)) %>%
  filter(count > 20, dist < 2000)
  
  # using SE, fc, no pipes
  flight_summary_fc <- fc(filter_, .dots = list('count > 20', 'dist < 2000'),
                             .data = fc(summarize_, .dots = list(count = 'n()',
                                                      dist='mean(distance, na.rm=TRUE)',
                                                      delay='mean(arr_delay, na.rm=TRUE)'),
                             .data = fc(group_by_, .dots = list('tailnum'))(.data))(.data))
  flight_summary_fc(flights)
  
  # using SE, fc, with pipes
  flight_summary_fc_pipe <- fc(group_by_, .dots = list('tailnum')) %>%
  fc(summarize_, .dots = list(count = 'n()',
                              dist='mean(distance, na.rm=TRUE)',
                              delay='mean(arr_delay, na.rm=TRUE)')) %>%
  fc(filter_, .dots = list('count > 20', 'dist < 2000'))
  flight_summary_fc_pipe(flights)
\end{example}

Additionally, we note that \verb|flight_summary_fc()| and \verb|flight_summary_fc_pipe()| can also operate on database objects hosted via \CRANpkg{dbplyr} \citep{dbplyr}. Specifically, instead of applying each function to an actual data frame, we could leverage the extended capabilities of \verb|filter_()|, \verb|group_by_()|, and \verb|summarize_()| to operate on a data frame object. 

The chunk of code below creates a database in memory from the \verb|flights| data frame. 

\begin{example}
  my_db <- DBI::dbConnect(RSQLite::SQLite(), path = ":memory:")
  copy_to(my_db,
          flights,
          temporary = FALSE,
          indexes = list(
            c("year", "month", "day"),
            "carrier", "tailnum"
           )
          )
 flights_db <- tbl(my_db, "flights")
\end{example}

\verb|flights_db| is now a database object and we could run \verb|flight_summary_fc(flights_db)|.

\subsection{Example 4} % (fold)
\label{sub:example_4}

This example illustrates how anonymous functions may be used in function composition with \pkg{fc}. Consider writing a function that (1) shuffles the rows of a data frame, (2) outputs the first ten rows and a subset of columns, and (3) prints a summary of each column. This could be written a number of ways using \pkg{fc}. We begin by examining two ways of composing functions without pipes.

\begin{example}
get_sepal1 <- fc(summary, object = fc(head, x = (function(x) {
                              x[sample(1:nrow(x)), 
                                grep("Sepal", colnames(x))]
                              }) (x), n = 10)(x))
get_sepal2 <- fc(summary, object = fc(head, x = fc(function(x, cols) {x[sample(1:nrow(x)), cols]},
                               cols = grep("Sepal", colnames(x)))(x), n = 10)(x))                             
\end{example}

In \verb|get_sepal1()|, we define an anonymous function whose result when run on the input variable \verb|x| will be passed along as the primary argument for \verb|head|. In \verb|get_sepal2()|, we use an anonymous function as the primary argument to \verb|fc| and use partial valuation to set the value of \verb|cols|. 

We can rewrite each of these function compositions with pipes:

\begin{example}
get_sepal1_pipe <- (function(x) {
  x[sample(1:nrow(x)), grep("Sepal", colnames(x))]
}) %>% fc(head, n=10) %>% summary
get_sepal2_pipe <- fc(function(x, cols) {x[sample(1:nrow(x)), cols]},
                               cols = grep("Sepal", colnames(x))) %>% 
                               fc(head, n = 10) %>% summary
\end{example}

\subsection{Comparison}

We now compare the run speeds of Examples 1 through 4 to comparable implementations achieved using \CRANpkg{magrittr}, \CRANpkg{purrr}, and base R. For each example, we present code for the various implementations. The suffixes of the composed functions indicate the implementation approach:
\begin{itemize}
  \item \texttt{\_base}: function is created from explicitly writing out the function composition in base R.
  \item \texttt{\_purrr}: function is created using \CRANpkg{purrr}'s \verb|compose()| function (and its \verb|partial()| function, where applicable).
  \item \texttt{\_mag}: function is created using \CRANpkg{magrittr}, with a \verb|.| in the first entry of the pipeline.
%   \item \texttt{\_fc}: function is composed using \pkg{fc}, with possible nesting.
%   \item \texttt{\_fc\_pipe}: function is composed using \pkg{fc} using pipes for readability.
\end{itemize}

\begin{example}
  ## Example 1
  log_sqrt_base <- function(x) log(x=sqrt(x))
  log_sqrt_purrr <- compose(log, sqrt)
  log_sqrt_mag <- . %>% sqrt %>% log

  ## Example 2
  search_trim_base <- function(v) {
    trimws(gsub(grep(v, pattern="<[^/]*>", value=TRUE), 
    pattern=".*>(.*)<.*", replacement = "\\1"))
  }
  search_trim_mag <- . %>% grep(pattern="<[^/]*>", x=., value=TRUE) %>%
    gsub(".*>(.*)<.*", "\\1", x=.) %>%
    trimws
  search_trim_purrr <- purrr::compose(trimws, partial(gsub, pattern=".*>(.*)<.*",   
                                        replacement = "\\1"), 
                                partial(grep, pattern="<[^/]*>", value=TRUE))

  ## Example 3
  flight_summary_base <- function(x) {
    filter_(summarize_(group_by_(x, .dots = list('tailnum')), 
                        .dots = list(count = 'n()',
                                     dist='mean(distance, na.rm=TRUE)',
                                     delay='mean(arr_delay, na.rm=TRUE)')),
              .dots = list('count > 20', 'dist < 2000'))
  }
  flight_summary_mag <- . %>% group_by_(.dots = list('tailnum')) %>%
                              summarize_(.dots = list(count = 'n()',
                                         dist='mean(distance, na.rm=TRUE)',
                                         delay='mean(arr_delay, na.rm=TRUE)')) %>%
                              filter_(.dots = list('count > 20', 'dist < 2000'))

  flight_summary_purrr <- compose(partial(filter_, .dots = list('count > 20', 'dist < 2000')), 
                               partial(summarize_, .dots = list(count = 'n()',
                                  dist='mean(distance, na.rm=TRUE)',
                                  delay='mean(arr_delay, na.rm=TRUE)')),
                                        partial(group_by_, .dots = list('tailnum')))
                                        
  ## Example 4
  get_random_sepal_base <- function(x) head(x[sample(1:nrow(x)), 
                                grep("Sepal", colnames(x))], n=10)
  get_random_sepal_mag <- . %>% (function(x) x[sample(1:nrow(x)), grep("Sepal", colnames(x))]) %>% 
                            head(n = 10) %>% 
                            summary
  get_random_sepal_purrr <- compose(function(x) {
                              x[sample(1:nrow(x)),
                                grep("Sepal", colnames(x))]
                            }, partial(head, n=10), summary)
\end{example}
Notes: In Example 3, we use the standard evaluation flavors of the \pkg{dplyr} functions rather than their NSE counterparts, even for use with \pkg{magrittr} so as to ensure a fair comparison; runtimes reflect those of the functions applied to the database object instead of the data frame. For Example 4, we previously showed multiple ways of composing functions to achieve the desired functionality via \pkg{fc}. For comparison purposes, we will use \verb|get_sepal1()| and \verb|get_sepal1_pipe()|.

Each function was run 10,000 times with the same inputs. Runtimes are checked using the \CRANpkg{microbenchmark} package \citep{microbenchmark}. The results are summarized below.

% latex table generated in R 3.3.2 by xtable 1.8-2 package
% Wed May 23 14:02:46 2018
\begin{table}[ht]
\centering
\begin{tabular}{rllll}
  \hline
 & Example 1 ($\mu s$) & Example 2 ($\mu s$) & Example 3 ($ms$) & Example 4 ($ms$) \\ 
  \hline
base R & $0.782 (\pm0.01)$ & $74.584 (\pm0.317)$ & $4.418 (\pm0.02)$ & $0.284 (\pm0.002)$ \\ 
  magrittr & $5.386 (\pm0.053)$ & $81.174 (\pm0.55)$ & $4.468 (\pm0.027)$ & $1.503 (\pm0.016)$ \\ 
  purrr & $5.486 (\pm0.17)$ & $82.399 (\pm0.498)$ & $4.51 (\pm0.027)$ & $2.681 (\pm0.022)$ \\ 
  fc & $0.87 (\pm0.034)$ & $76.976 (\pm0.434)$ & $4.445 (\pm0.023)$ & $1.498 (\pm0.007)$ \\ 
  fc\_pipe & $2.537 (\pm0.159)$ & $77.553 (\pm0.448)$ & $4.431 (\pm0.014)$ & $1.499 (\pm0.007)$ \\ 
   \hline
\end{tabular}
\caption{Runtime comparisons for Examples 1 through 4, showing mean runtime ($\pm$ standard error) across 10,000 iterations. Examples 1 and 2 runtimes are reported in units of microseconds whereas Example 3 and 4 runtimes are reported in milliseconds.}
\end{table}

Examples 1 and 2 show \pkg{fc}'s superior performance when used with standard R functions. We hypothesize that \pkg{fc} does especially well with Example 1 because the constituent functions of \code{log()} and \code{sqrt()} are simple enough that most of the heavy-lifting done by \pkg{magrittr} and \pkg{purrr} involve parsing their NSE expressions. For that reason, when pipes are used with \pkg{fc}, the runtime suffers somewhat as well, albeit to a lesser extent. In Example 2, while the constituent functions are not primitive, like those of Example 1, their specified arguments involve expressions that  require a fair amount of parsing. 

In Example 3, had we used the NSE versions of the \pkg{dplyr} functions like \verb|filter()|, \verb|group_by()|, and \verb|summarize()| for the \pkg{magrittr} solution, we would find that \pkg{magrittr}'s solution runs much faster. This is due to clever implementation in the NSE \pkg{dplyr} functions themselves (e.g. \verb|filter()| simply works faster than \verb|filter_()|) and not due to differences between function composition between \pkg{magrittr} and \pkg{fc}. 

Example 4 compares how the various approaches fare with anonymous functions used in composition. It seems as though \pkg{magrittr} does about as well as \pkg{fc} on this front. 

Overall, from the perspective of runtime, we find that \pkg{purrr} usually does as well as \pkg{magrittr}, but could at times do worse. \pkg{fc} typically does as well as \pkg{magrittr} but could be much faster. 

\section{Discussion} 
\label{Discussion}

%The goal of this project is to identify and implement key functionality provided by \pkg{magrittr} using standard evaluation provided in the core R language. The goal is not to subvert or compete with \pkg{magrittr}, which has already found widespread adoption.
In this article, we presented a new package \pkg{fc} that combines general function composition with the economy and clarity afforded by a pipeline syntax. Importantly, \pkg{fc} utilizes standard evaluation so that components of a pipeline can be directly evaluated in R. This is in stark contrast to the \pkg{magrittr} package, which (1) utilizes a pipeline syntax to chain functions without composition and (2) uses non-standard evaluation to further prioritize conciseness of syntax. 
%This exercise has shown that a large portion of the qualitative value provided by \code{magrittr} is the economy and clarity with which sequences of operations can be defined.
%Compared to \pkg{fc}'s usage, \pkg{magrittr}'s uses fewer characters and can be easier to read. However, by not compromising on the use of standard evaluation, functions created via \pkg{fc} can be more performant when a pipeline is deep or is called many times. 

\pkg{fc}'s usage can be slightly more verbose compared to that of \pkg{magrittr}, since expressions in the \pkg{fc} pipeline are evaluated without manipulation. However, the return of the pipe is a human-readable function that is consistent with expectations given a sequence of functions defined in the pipe. Furthermore, because the result is a composed function, rather than a sequence of functions, the resulting syntax tree is easier to debug and optimize. As a consequence, \pkg{fc}'s composition provides similar performance as standard R implementations and tends to achieve better runtimes than \CRANpkg{magrittr} and \CRANpkg{purrr}, particularly when a pipeline is deep or is called many times. 

%Because of its better performance along with the fact that returned functions are composed R functions, \pkg{fc} may be useful for constructing functions created in \pkg{magrittr}. The only incompatibility would be that the current implementation of \pkg{fc} does not support expressions as function argument names, which is used in functions including \code{subset()}. However, this could be addressed by either, detecting expressions as arguments and not optimizing in that case, or by adding this NSE capability.

\bibliography{RJreferences}

\address{Xiaofei (Susan) Wang, PhD\\
  Department of Statistics and Data Science, Yale University\\
  24 Hillhouse Ave.\\
  New Haven, CT, 06520\\
  USA\\
  ORCiD: 0000-0003-0532-930X\\
  \email{xiaofei.wang@yale.edu}}

\address{Michael John Kane, PhD\\
  Yale School of Public Health Department of Biostatistics\\
  60 College Street\\
  New Haven, CT, 06510\\
  USA\\
  ORCiD: 0000-0003-1899-6662\\
  \email{michael.kane@yale.edu}}
\end{article}

\end{document}